\documentclass[aps,twocolumn,amsmath,amssymb,superscriptaddress,preprintnumbers]{revtex4-1}
\usepackage[pdftex]{graphicx}
\usepackage[outdir=./]{epstopdf}
\usepackage{enumerate,ulem}
\newcommand \beq{\begin{eqnarray}}
\newcommand \eeq{\end{eqnarray}}

\newcommand{\bfr}{\boldsymbol{r}}
\newcommand{\bfk}{\boldsymbol{k}}

\begin{document}

\title{Strain-induced nonlinear spin Hall effect in topological Dirac semimetal}
\author{Yasufumi Araki}
\affiliation{Institute for Materials Research, Tohoku University, Sendai 980-8577, Japan}
\affiliation{Frontier Research Institute for Interdisciplinary Sciences, Tohoku University, Sendai 980-8578, Japan}

\begin{abstract}
We show that an electric field applied to a strained topological Dirac semimetal,
such as $\mathrm{Na_3 Bi}$ and $\mathrm{Cd_3 As_2}$,
induces a spin Hall current that is quadratic in the electric field.
By regarding the strain as an effective ``axial magnetic field'' for the Dirac electrons,
we investigate the electron and spin transport semiclassically in terms of the chiral kinetic theory.
The nonlinear spin Hall effect arises as the cross effect between
the regular Hall effect driven by the axial magnetic field
and the anomalous Hall effect coming from the momentum-space topology.
It provides an efficient way
to generate a fully spin-polarized and rectified spin current out of an alternating electric field,
which is sufficiently large and can be directly tuned by the gate voltage and the strain.
\end{abstract}


\maketitle

The idea of spin current has significantly developed the field of nanoscale condensed matter physics,
in particular of spintronics \cite{Zutic_2004,Maekawa_spin_current,Takahashi_2008}.
Spin current plays an important role in controlling and detecting magnetization in magnetic nanostructures.
The spin Hall effect (SHE) is one of the ways to obtain a spin current,
in particular a pure spin current transverse to the injected charge current \cite{Jungwirth_2012,Hoffmann_2013,Sinova_2015}.
Its reciprocal effect, namely the inverse SHE (ISHE),
converts the injected spin current into a charge current, which is useful in detecting a pure spin current \cite{Saitoh_2006,Valenzuela_2006,Zhao_2006}.
The SHE is efficient in that it does not require any ferromagnetic material,
which makes the system free from stray magnetic field.

The origin of the SHE can be classified into the extrinsic and intrinsic mechanisms.
While the extrinsic mechanism is triggered by spin-asymmetric scattering
at impurities with spin-orbit coupling (SOC) \cite{Dyakonov_Perel,Hirsch_1999,Zhang_2000},
the intrinsic effect originates from the nontrivial band topology due to SOC \cite{Murakami_2003,Sinova_2004}.
Since SOC violates the spin conservation,
the spin current generated by the intrinsic SHE usually gets suppressed as it flows by a long distance.
However, in some topological materials such as HgTe quantum well,
the spin-orbit field (approximately) preserves U(1) spin symmetry by a certain quantization axis (e.g. $S_z$),
yielding a spin Hall current that is fully polarized along the quantization axis.
Its spin Hall conductivity is quantized,
which is related to the $\mathbb{Z}_2$ topology of the eigenstate \cite{Murakami_2004,Kane_2005,Bernevig_2006}.

In three dimensions (3D), topological Dirac semimetals (TDSMs) \cite{Wang_2012,Wang_2013},
such as $\mathrm{Na_3 Bi}$ \cite{Liu_2014} and $\mathrm{Cd_3 As_2}$ \cite{Neupane_2014},
show the intrinsic SHE protected by $\mathbb{Z}_2$ topology.
TDSMs are characterized by pair(s) of Dirac points (DPs/valleys) separated in momentum space,
which are protected by rotational symmetry around an axis.
The intrinsic spin Hall conductivity is determined by the separation of the DPs in momentum space \cite{Yang_2014,Yang_2015,Gorbar_2015,Burkov_2016},
which is analogous to the anomalous Hall effect (AHE) in a Weyl semimetal (WSM) with broken time-reversal symmetry (TRS) \cite{Burkov_2011,Yang_2011}.
The intrinsic SHE in TDSM is thus robust against disorders in bulk,
and the value of spin Hall conductivity is fixed for each material.

\begin{figure}[bp]
\begin{tabular}{ccc}
\includegraphics[height=4cm]{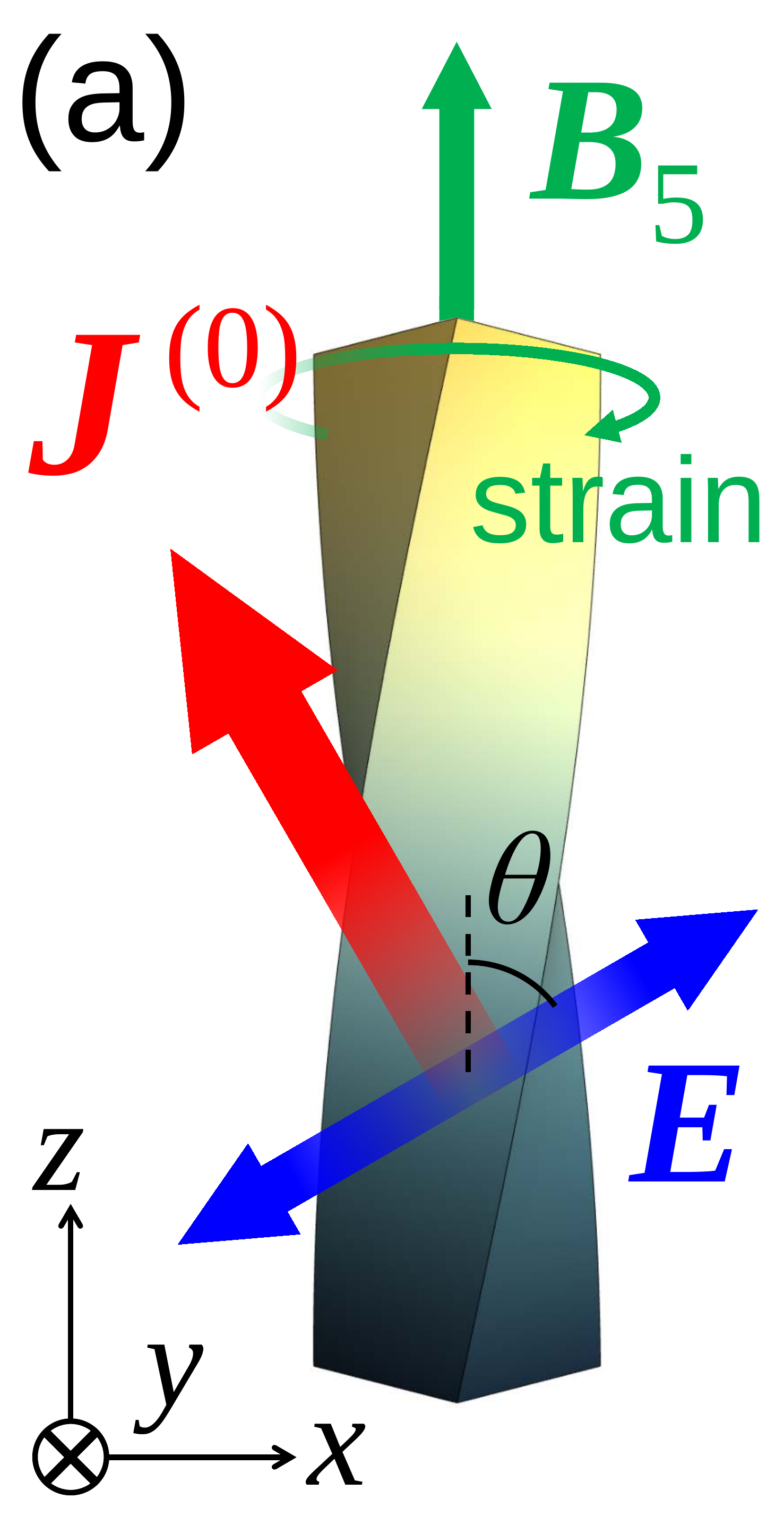} & \hspace{0.1cm} & \includegraphics[height=4cm]{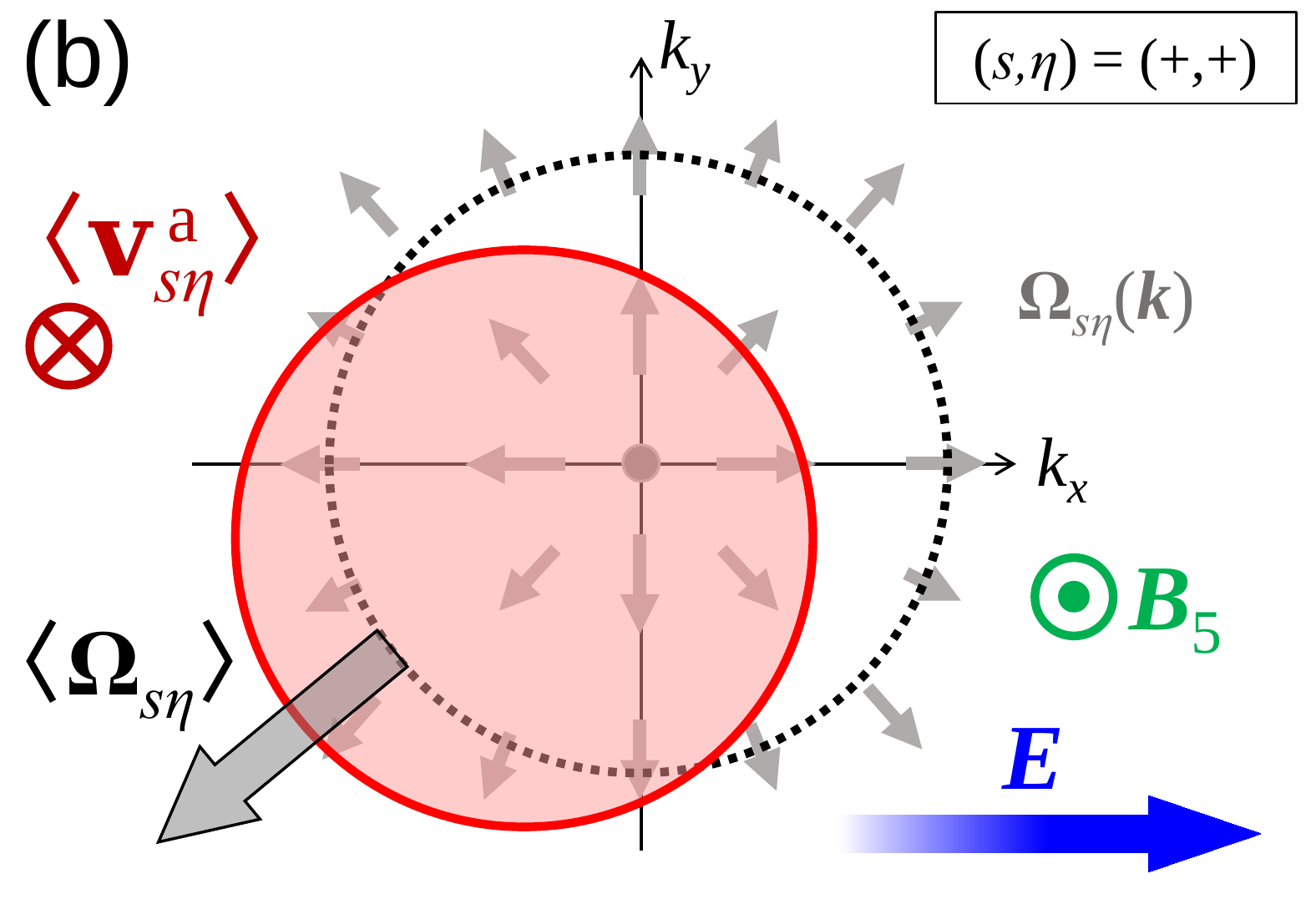}
\end{tabular}
\caption{
\textbf{Schematic pictures for the nonlinear spin Hall effect in topological Dirac semimetal.}
(a) The setup of the system.
A lattice strain on the topological Dirac semimetal (TDSM) is equivalent to the axial magnetic field $\boldsymbol{B}_5$.
An alternating electric field $\boldsymbol{E}$ drives a rectified spin current $\boldsymbol{J}^{(0)}$ quadratic in $\boldsymbol{E}$.
(b) The electron distribution in momentum space in response to the electric field $\boldsymbol{E}$ and the axial magnetic field $\boldsymbol{B}_5$.
The distribution is shifted from the equilibrium distribution (dashed circle) transverse as well as longitudinal to $\boldsymbol{E}$ at linear response (red solid circle),
due to the regular Hall effect (RHE) under $\boldsymbol{B}_5$.
It induces an imbalance in the Berry curvature $\boldsymbol{\Omega}_{s\eta}$ (small gray arrows),
which leads to the anomalous velocity $\boldsymbol{v}^\mathrm{a}_{s\eta}$ as the second-order response in $\boldsymbol{E}$.}
\label{fig:distribution}
\end{figure}

Therefore, in order to tune and enhance the spin Hall current from its fixed value in TDSM,
we need to go beyond the linear response regime with respect to the electric field,
which is neccessary in making use of TDSM as an efficient spin current injector.
Nonlinear spin current generation is important for device application
in that it generates a rectified (stationary) spin current from an alternating electric field, or a light,
which has been proposed in transition metal dichalcogenides \cite{Yu_2014} and 2D Rashba--Dresselhaus systems \cite{Hamamoto_2017}.
Moreover, the nonlinear transport is important also from the topological point of view;
a recent study has shown that the momentum-space Berry curvature gives rise to the nonlinear Hall transport \cite{Sodemann_2015}.
For Dirac/Weyl semimetals, in particular, nonlinear charge current generation is proposed in several hypothetical setups,
in which the strong Berry flux around the Dirac/Weyl points gives rise to the nonlinear current \cite{Morimoto_2016,Zyuzin_2017}.
Nonlinear spin current generation might be of equal significance in Dirac/Weyl semimetals,
although it has not been taken into account so far.

In this work,
we demonstrate the nonlinear (quadratic) SHE in TDSM
by introducing a lattice strain to the system.
A lattice strain on a TDSM effectively serves as a valley-dependent magnetic field,
namely the axial magnetic field \cite{Cortijo_2015,Pikulin_2016,Liu_2017,Grushin_2016},
which is essential here in filtering the spin and valley degrees of freedom [see Fig.~\ref{fig:distribution}(a)].
We make use of the chiral kinetic theory,
which describes the dynamics and distribution of the Dirac electrons for each valley \cite{Stephanov_2012,Son_Yamamoto_2013,Son_Spivak_2013,Gorbar_2017a,Zyuzin_V},
and derive the spin Hall current semiclassically up to the second order in the electric field.
This nonlinear SHE can be regarded as the cross effect between the regular Hall effect (RHE) induced by the axial magnetic field
and the AHE induced by the momentum-space topology \cite{Karplus_Luttinger,Jungwirth_2002,Onoda_2002}:
the external electric field together with the axial magnetic field
shifts the electron distribution in momentum space by the Lorentz force,
and this shifted distribution yields the anomalous velocity due to the momentum-space Berry curvature,
leading to the spin Hall current in total [see Fig.~\ref{fig:distribution}(b)].
We find that the nonlinear spin Hall current can be tuned by the gate voltage (electron chemical potential),
and can reach the value comparable to the linear intrinsic spin Hall current,
at the electric field $\sim 10 \mathrm{kV/m}$.
The spin current generated by this effect is fully spin-polarized and rectified
even though the driving electric field is alternating,
which we expect to be useful in designing TDSM-based spintronic devices.

\

\section*{Results}

\textbf{Topological Dirac semimetal and strain.} ---
We start with the low-energy effective Hamiltonian for TDSM,
\begin{equation}
H(\bfk) = v_\mathrm{F} \left[\sigma_z \tau_x k_x - \tau_y k_y + \eta \tau_z (k_z - \eta k_\mathrm{D}) \right], \label{eq:Hamiltonian}
\end{equation}
with $v_\mathrm{F}$ the material-dependent Fermi velocity \cite{Wang_2012,Wang_2013}.
This minimal Hamiltonian consists of the atomic orbital degrees of freedom
(e.g. Na-$3s$ and Bi-$6p$ for $\mathrm{Na_3 Bi}$) labeled by the Pauli matrix $\boldsymbol{\tau}$
and the spin (up/down) degrees of freedom labeled by $\boldsymbol{\sigma}$.
The Hamiltonian is linearized around the two DPs,
which reside at $\bfk = (0,0, \eta k_\mathrm{D})$ with $\eta = \pm$ respectively.
Each DP is doubly degenerate and is protected by the crystalline rotational symmetry around $z$-axis \cite{Yang_2014}.
In the vicinity of the DPs,
the energy eigenvaule for the electron (conduction) band is given as $\epsilon(\bfk) = v_\mathrm{F} |\bfk - \eta k_\mathrm{D}  \hat{\boldsymbol{e}}_z|$.

In the absence of nonlinear corrections in $\bfk$,
$\sigma_z$ behaves as a good quantum number,
which we denote $s=\pm$ or spin up/down.
For each $s=\pm$, the Hamiltonian takes the same form as that for a WSM with broken TRS.
The topological charge for the valley $\eta$ with spin $s$ is $\nu_{s\eta} = s\eta$;
the net topological charge cancels within each valley and within each spin \cite{Yang_2015}.
This system shows the intrinsic SHE linear in the electric field,
protected by the $\mathbb{Z}_2$ topology,
with the spin Hall conductivity $\sigma_{xy}^\mathrm{S} = (e^2/\pi^2)k_\mathrm{D}$ \cite{Burkov_2016}.

A lattice strain modifies this Hamiltonian,
by generating additional hopping terms that arise from the lattice dislocation.
This effect can be effectively described as an \textit{axial magnetic field} $\boldsymbol{B}_5$ in the vicinity of the DPs,
which is almost equivalent to the normal magnetic field
but couples to each valley with the opposite sign $(\eta)$ \cite{Cortijo_2015,Pikulin_2016,Liu_2017,Grushin_2016}.
Such a correspondence is known in various crystalline systems such as graphene \cite{Gonzalez_1992,Suzuura_2002,Guinea_2010}.
In TDSMs, such as $\mathrm{Na_3 Bi}$ and $\mathrm{Cd_3 As_2}$,
a screw strain on a nanowire can generate $\boldsymbol{B}_5$ up to $0.3 \mathrm{T}$ \cite{Pikulin_2016},
and a bending of a thin film can make it up to $15 \mathrm{T}$ \cite{Liu_2017},
which are large enough to reproduce the Landau quantization.
In this work, we assume that $\boldsymbol{B}_5$ is macroscopically uniform for simplicity,
and investigate the electron and spin transport up to the linear order in $\boldsymbol{B}_5$.

\textbf{Field-induced current.} ---
In the present work,
we focus on the electron transport driven by the alternating electric field with the frequency $\omega$,
defined by $\boldsymbol{E}(t) = 2\boldsymbol{E}_0 \cos\omega t$,
which can account for a linearly polarized light as well.
We omit the real magnetic field $\boldsymbol{B}$,
whereas fix the strain-induced axial magnetic field $\boldsymbol{B}_5$
finite and (locally) homogeneous.
By solving the Boltzmann equation for the electrons in terms of the chiral kinetic theory (see Methods),
we estimate the driven current $\boldsymbol{j}_{s\eta}(t)$ for each spin $s$ and valley $\eta$
up to the first order in $\boldsymbol{B}_5$ and the second order in $\boldsymbol{E}_0$.
While the linear response to the electric field $\boldsymbol{E}$ yields an alternating current $\boldsymbol{j}_{s\eta}^{(\pm \omega)}$,
the quadratic response consists of the second harmonic part $\boldsymbol{j}_{s\eta}^{(\pm 2\omega)}$ and the stationary (rectified) part $\boldsymbol{j}_{s\eta}^{(0)}$,
where the superscript with $(\cdot)$ on a physical quantity denotes its oscillation frequency.

Up to quadratic response to the electric field,
we find that the stationary part $\boldsymbol{j}_{s\eta}^{(0)}$ depends only on the spin $s$ but not on the valley $\eta$,
namely $\boldsymbol{j}_{s\eta}^{(0)} \equiv (s/4) \boldsymbol{J}^{(0)}$.
As a result, we obtain no net charge current but a pure spin current $\boldsymbol{J}^{(0)}$,
with its quantization axis taken to $S_z$.
This stationary spin current consists of the equilibrium part $\boldsymbol{J}^{(0)}_{\mathrm{eq}}$ that is independent of the electric field $\boldsymbol{E}_0$
 and the nonequilibrium part $\boldsymbol{J}^{(0)}_{\mathrm{neq}}$ that is quadratic in $\boldsymbol{E}_0$.
The equilibrium spin current
\begin{equation}
\boldsymbol{J}^{(0)}_{\mathrm{eq}} = -\frac{e^2}{\pi^2} \mu \boldsymbol{B}_5 \label{eq:J-eq}
\end{equation}
is the axial counterpart of the chiral magnetic effect,
sometimes referred to as the \textit{chiral axial magnetic} or \textit{chiral pseudomagnetic effect} \cite{Zhou_2012,Huang_2017,Gorbar_2017b}.
It comes from all the occupied states below the Fermi level,
which is robust against disorder but cannot be taken out of the sample.
On the other hand, the nonequilibrium part is given as
\begin{equation}
\boldsymbol{J}^{(0)}_{\mathrm{neq}} = - \frac{4 e^2 v_\mathrm{F}^2}{3\pi^2 \mu} \frac{\tau^2}{(1+\omega^2\tau^2)^2} (\boldsymbol{B}_5 \times \boldsymbol{E}_0) \times \boldsymbol{E}_0, \label{eq:J-neq}
\end{equation}
where $\tau$ is the relaxation time for all the relaxation processes,
including the intravalley, intervalley, and spin-flip processes.
This nonequilibrium spin current is carried by the electrons at the Fermi surface,
and can be extracted out of the sample.
Since this is the \textit{spin} current that flows \textit{perpendicular} to the electric field $\boldsymbol{E}_0$
and is \textit{quadratic} in $\boldsymbol{E}_0$,
we may call this effect the nonlinear spin Hall effect.

\textbf{Origin of the nonlinear spin Hall effect.} ---
This nonlinear spin Hall current can be regarded as the interplay effect between the regular Hall effect (RHE) and the anomalous Hall effect (AHE) as follows:
Figure \ref{fig:distribution}(b) shows its schematic picture.
At the first order in the electric field,
the Lorentz force by the axial magnetic field
shifts the distribution $f_{s\eta}(\bfk)$ for each spin $s$ and valley $\eta$
to the direction of $-\eta (\boldsymbol{B}_5\times \boldsymbol{E})$,
which accounts for the RHE.
For each $\bfk$ in this shifted distribution,
the anomalous velocity, which accounts for the intrinsic AHE in various TRS-broken systems,
is given as $\boldsymbol{v}^\mathrm{a}_{s\eta} \sim \boldsymbol{E} \times \boldsymbol{\Omega}_{s\eta} \sim s\eta \boldsymbol{E} \times \hat{\bfk}$,
using the $\bfk$-space Berry curvature $\boldsymbol{\Omega}_{s\eta}(\bfk) = s\eta \bfk/2k^3$ around each Dirac point.
Integrating the anomalous velocity over the whole $\bfk$-space,
its contribution to the current can be qualitatively estimated as
\begin{eqnarray}
\boldsymbol{j}^\mathrm{a}_{s\eta} &=& -e\int \frac{d^3\bfk}{(2\pi)^3} \ \boldsymbol{v}^\mathrm{a}_{s\eta}(\bfk) f_{s\eta}(\bfk) \\
 &\sim& -s\eta \boldsymbol{E} \times \int d^3\bfk \ \hat{\bfk} f_{s\eta}(\bfk) \sim s\boldsymbol{E} \times (\boldsymbol{B}_5\times \boldsymbol{E}), \nonumber
\end{eqnarray}
which accounts for the nonlinear spin Hall current given in Eq.~(\ref{eq:J-neq}).
In this sense, we can regard the nonlinear SHE found here as the combination of the RHE and the AHE,
or the interplay between the real-space topology and the momentum-space counterpart.
[The Lorentz force for the RHE is imprinted in the second term in Eq.~(\ref{eq:eom-k}),
while the anomalous velocity for the AHE appears in the second term in Eq.~(\ref{eq:eom-r});
see Methods for details.]

\textbf{How to detect the nonlinear spin Hall current.} ---
We are curious if the nonlinear spin Hall current obtained above can be observed experimentally.
First, we estimate the typical magnitude of this spin current $\boldsymbol{J}^{(0)}_{\mathrm{neq}}$,
by comparing it with other major spin currents,
namely the equilibrium spin current $\boldsymbol{J}^{(0)}_{\mathrm{eq}}$ given by Eq.~(\ref{eq:J-eq}),
and the linear intrinsic spin Hall current $\boldsymbol{J}^{(\pm\omega)}_{\mathrm{int}} = \sigma_{xy}^\mathrm{S} (\hat{\boldsymbol{e}}_z \times \boldsymbol{E}_0)$.
As we have mentioned above, we explicitly supplement the linear intrinsic spin Hall current here,
which is not included in the present chiral kinetic theory analysis.
Although $\boldsymbol{J}^{(\pm\omega)}_{\mathrm{int}}$ driven by the AC electric field $\boldsymbol{E}(t)$ is alternating with the frequency $\omega$,
we shall compare it with the stationary spin currents to see which effect is the most dominant.

Using Eqs.~(\ref{eq:J-eq}) and (\ref{eq:J-neq}),
the ratios among $\boldsymbol{J}^{(0)}_{\mathrm{neq}}$, $\boldsymbol{J}^{(0)}_{\mathrm{eq}}$,
and $\boldsymbol{J}^{(\pm\omega)}_{\mathrm{int}}$ are given as
\begin{equation}
\frac{J^{(0)}_{\mathrm{neq}}}{J^{(0)}_{\mathrm{eq}}} = \frac{4}{3} \left( \frac{e E_0 v_\mathrm{F} \tau}{\mu Z_\omega} \right)^2, \ 
\frac{J^{(0)}_{\mathrm{neq}}}{J^{(\pm\omega)}_{\mathrm{int}}} = \frac{4}{3} \frac{E_0 B_5}{\mu k_\mathrm{D}}\left(\frac{e v_\mathrm{F} \tau}{Z_\omega}\right)^2,
\end{equation}
where $Z_\omega = 1+\omega^2 \tau^2$.
Here we employ the material parameters
$v_\mathrm{F} = 0.5 \times 10^6 \mathrm{m/s}$ and $k_\mathrm{D} = 0.95 \mathrm{nm^{-1}}$ observed in $\mathrm{Na_3 Bi}$ \cite{Liu_2014},
and use the typical values $\mu = 10 \mathrm{meV}$ and $\tau = 1\mathrm{ps}$.
If an electric field $E_0 = 10^4 \mathrm{V/m}$
alternating in frequency $\omega \ll \tau^{-1}$ is applied to a bulk TDSM
under a strain equivalent to the axial magnetic field $B_5 = 1\mathrm{T}$,
the ratios among the induced currents are estimated as $J^{(0)}_{\mathrm{neq}}/J^{(0)}_{\mathrm{eq}} = 0.33$ and $J^{(0)}_{\mathrm{neq}}/J^{(\pm\omega)}_{\mathrm{int}} = 0.5$.
From these ratios,
we find that the nonlinear spin Hall current becomes comparable to the other two equilibrium spin currents
under typical strengths of fields,
which implies that the nonlinear spin Hall current is significant enough to be experimentally measured.

Next, let us check the orientation of the nonlinear spin Hall current
and discuss how it can be detected experimentally.
We define $z$-axis as the center of strain, i.e. $\boldsymbol{B}_5 = B_5 \hat{\boldsymbol{e}}_z$,
and introduce the electric field $\boldsymbol{E}_0$ tilted from $\boldsymbol{B}_5$ by the angle $\theta$,
i.e. $\boldsymbol{E}_0 = E_0(\cos\theta \hat{\boldsymbol{e}}_z + \sin\theta \hat{\boldsymbol{e}}_x)$ [see Fig.~\ref{fig:distribution}(a)].
Then the nonlinear spin Hall current $\boldsymbol{J}^{(0) \mathrm{neq}}$ flows in parallel to
\begin{equation}
-(\boldsymbol{B}_5 \times \boldsymbol{E}_0) \times \boldsymbol{E}_0 = B_5 E_0^2 \sin\theta \left( \sin\theta \hat{\boldsymbol{e}}_z - \cos\theta \hat{\boldsymbol{e}}_x \right). \label{eq:angular-dependence}
\end{equation}
As we can easily see from this equation,
the nonlinear spin Hall current vanishes when $\boldsymbol{E}_0 \parallel \boldsymbol{B}_5$ (i.e. $\theta=0,\pi$).
On the other hand,
it is maximized when $\boldsymbol{E}_0 \perp \boldsymbol{B}_5$ (i.e. $\theta=\pi/2$),
flowing in parallel to $\boldsymbol{B}_5$ ($z$-direction).
If $\boldsymbol{E}_0$ is at the intermediate angle, the spin current flows in $x$-direction as well as $z$-direction.

The detection method of the spin current depends on its direction.
The $z$-component of the spin current, flowing parallel to the screw strain axis,
can be easily extracted from the system by putting a spin-sensitive material at the end of this axis.
One can make use of a ferromagnetic metal or semiconductor,
in which the injected spin current invokes a spin torque on the magnetization,
leading to an oscillation or a switching of the magnetization.
Heavy metals such as Pt can also be used,
in which the spin current is converted to a charge current via the ISHE.
On the other hand,
the $x$-component of the spin current can be measured without any such external probes:
the spin current flowing in $x$-direction can induce a charge current in $y$-direction
via the (intrinsic) ISHE in the TDSM itself.
Using the spin Hall angle $\theta_\mathrm{SH} = \tilde{\sigma}_{xy} / \sigma_{xx}$,
with $\sigma_{xx}$ the in-plane longitudinal conductivity of the TDSM,
the induced charge current can be given as $\boldsymbol{j}^{(0)}_\mathrm{ISH} = \theta_\mathrm{SH} \hat{\boldsymbol{e}}_z \times \boldsymbol{J}^{(0)}_{\mathrm{neq}}$.
The $\theta$-dependence shown in Eq.~(\ref{eq:angular-dependence}) may be checked by these measurements,
with sweeping the direction of the $\boldsymbol{E}$-field.

\

\section*{Discussion}

In this work, we have focused on a strained TDSM (e.g. $\mathrm{Na_3 Bi}$, $\mathrm{Cd_3 As_2}$, etc.),
and have shown that such a system shows a significant nonlinear SHE,
i.e. an external electric field induces a spin current perpendicular to the electric field
as its quadratic response.
This effect is described effectively by regarding the strain as the axial magnetic field $\boldsymbol{B}_5$,
namely the valley-dependent magnetic field.
The electron transport has been analyzed semiclasically in terms of the chiral kinetic theory.
The nonlinear SHE can be understood as the interplay effect between
the RHE due to the axial magnetic field $\boldsymbol{B}_5$
and the AHE due to the finite Berry curvature in momentum space.
This spin current reaches the magnitude comparable to the intrinsic spin Hall current
under the electric field $\sim 10^4 \mathrm{V/m}$,
and can be successfully tuned via the gate voltage (electron chemical potential) and the strain (axial magnetic field).
Our finding thus provides an efficient way to generate a rectified spin current out of an alternating electric field,
which may be useful for spin injection in future spintronic devices.

We have so far treated the disorder effect in terms of a single relaxation time $\tau$ for simplicity.
However, in a realistic TDSM,
the intravalley, intervalley, and spin-flipping scattering processes should be chracterized by distinct relaxation times.
In particular,
it is known that the $O(k^3)$ terms that become significant away from the DPs
violate the conservation of spin $S_z$,
which give rise to the spin-flip process in the presence of strong scatterers.
We leave the microscopic treatment of such scattering processes as an open question here.

As we have mentioned in the beginning,
since there is no term that violates the spin symmetry by $S_z$ around the DPs,
each spin block (up/down) of the topological Dirac Hamiltonian can be regarded
as the Weyl Hamiltonian with broken TRS.
Extracting a single spin block out of our analysis,
it can also account for the transport in TRS-broken WSMs.
In particular, in magnetic WSMs (e.g. $\mathrm{Mn_3 Sn}$),
an axial magnetic flux resides at a magnetic texture,
such as magnetic domain walls, vortices, skyrmions, etc. \cite{Liu_2013},
and its effect on the electronic spectrum has been verified both analytically and numerically \cite{Pikulin_2016,Grushin_2016,Araki_DW}.
In the presence of such an axial magnetic field,
our analysis implies that there arises the nonlinear Hall effect, inducing a charge current.
While the general theory of intrinsic nonlinear Hall effect was established
in terms of the momentum-space Berry curvature in the recent literature \cite{Sodemann_2015},
our setup also involves the real-space Berry curvature (axial magnetic field),
to which their theory cannot be applied as it is.
It will be another open question to find such theory with the Berry curvature involving the global phase space.

\

\section*{Methods}

\textbf{Chiral kinetic theory.} ---
In order to deal with the electron transport driven by the normal and axial electromagnetic fields,
we first need to understand the dynamics of an electron wave packet.
The dynamics of its center-of-mass position $\bfr$
and its gauge-invariant momentum $\bfk$ measured from the DP \cite{note1}
is described by the semiclassical equations of motion \cite{Chang_1995,Chang_1996,Sundaram_1999,Xiao_2010},
\begin{eqnarray}
\dot{\bfr} &=& \boldsymbol{\nabla}_{\bfk} \tilde{\epsilon}_{s\eta}(\bfk) - \dot{\bfk} \times \boldsymbol{\Omega}_{s\eta}(\bfk) \label{eq:eom-r} \\
\dot{\bfk} &=& -e\boldsymbol{E}_\eta - e\dot{\bfr} \times \boldsymbol{B}_{\eta} \label{eq:eom-k}
\end{eqnarray}
for each spin $s=\pm$ and valley $\eta =\pm$.
Here $\boldsymbol{E}_\eta = \boldsymbol{E}+\eta \boldsymbol{E}_5$ and $\boldsymbol{B}_\eta = \boldsymbol{B}+\eta\boldsymbol{B}_5$
denote the \textit{effective} electromagnetic fields for each valley.
Under the alternating electric field $\boldsymbol{E}(t) = 2\boldsymbol{E}_0 \cos \omega t$
and the lattice strain equivalent to the axial magnetic field $\boldsymbol{B}_5$,
the effective electromagnetic fields are given by
\begin{equation}
\boldsymbol{E}_\eta(t) = \boldsymbol{E}_0 (e^{i\omega t} + e^{-i\omega t}), \quad
\boldsymbol{B}_\eta = \eta \boldsymbol{B}_5.
\end{equation}
One should note that there are several modifications from the classical (Newtonian) equation of motion:
the electron energy is modified from its band dispersion $\epsilon(\bfk)$ by the orbital magnetic moment $\boldsymbol{m}_{s\eta}(\bfk)$ as
$\tilde{\epsilon}_{s\eta}(\bfk) = \epsilon(\bfk) - \boldsymbol{m}_{s\eta}(\bfk) \cdot \boldsymbol{B}_\eta$.
The momentum-space Berry curvature $\boldsymbol{\Omega}_{s\eta}(\bfk)$ gives rise to
the anomalous velocity $- \dot{\bfk} \times \boldsymbol{\Omega}_{s\eta}$,
which is the momentum-space counterpart of the Lorentz force $- e\dot{\bfr} \times \boldsymbol{B}_{\eta}$.
%
In the electron band of the TDSM, i.e. for $\epsilon(\bfk) = v_\mathrm{F} k$,
the quantities mentioned above are given as
\begin{equation}
\boldsymbol{m}_{s\eta}(\bfk) = s\eta \frac{e v_\mathrm{F}}{2k} \hat{\bfk}, \quad
\boldsymbol{\Omega}_\eta(\bfk) = s\eta \frac{1}{2k^2} \hat{\bfk}. \label{eq:Berry-curvature}
\end{equation}
Since both of them are significant in the vicinity of the DPs,
the nonlinear SHE discussed in this paper, which arises from these modifications,
becomes stronger at lower Fermi level.
%

Based on the single-particle dynamics discussed above,
we can describe the collective semiclassical dynamics of the electrons by the Boltzmann equation,
\begin{equation}
\left[ \dot{\bfr}\cdot\boldsymbol{\nabla}_{\bfr} + \dot{\bfk}\cdot\boldsymbol{\nabla}_{\bfk} + \partial_t \right]f_{s\eta}(\bfr,\bfk,t) = \left(\frac{df_{s\eta}}{dt}\right)_\mathrm{coll} \label{eq:Boltzmann}
\end{equation}
for the electron distribution function $f_{s\eta}(\bfr,\bfk,t)$ for each spin $s$ and valley $\eta$.
The collision term $(df_{s\eta}/dt)_\mathrm{coll}$ consists of various scattering processes contributing to relaxation;
here we approximate
$(df_{s\eta}/dt)_\mathrm{coll} = -[f_{s\eta}(\bfr,\bfk,t) - f_{s\eta}^\mathrm{eq}(\bfk)]/\tau$
with a single relaxation time $\tau$ for simplicity,
with which we incorporate spin relaxation and intervalley scattering processes as well as the intravalley process \cite{Son_Spivak_2013}.
$f_{s\eta}^\mathrm{eq}(\bfk) \equiv f^\mathrm{eq}(\tilde{\epsilon}_{s\eta}(\bfk))$ is the equilibrium distribution modified by the orbital magnetization.
Here we work with the chemical potential $\mu >0$ in the zero-temperature limit,
which gives $f^\mathrm{eq}(\epsilon) = \theta(\mu - \epsilon)$.
We here require the spatial homogeneity of the system,
so that the $\bfr$-dependence in $f_{s\eta}$ can be neglected.

By solving the kinetic equations [Eqs.~(\ref{eq:eom-r}) and (\ref{eq:eom-k})] and the Boltzmann equation [Eq.~(\ref{eq:Boltzmann})],
the current for each spin and valley can be evaluated by
\begin{equation}
\boldsymbol{j}_{s\eta}(t) = -e \int \frac{d^3 \bfk}{(2\pi)^3} D_{s\eta}(\bfk) \dot{\bfr} f_{s\eta}(\bfk,t), \label{eq:current}
\end{equation}
where $\dot{\bfr}$ is given as a function of $\bfk$ for each $s$ and $\eta$ by the solution of Eqs.~(\ref{eq:eom-r}) and (\ref{eq:eom-k}),
and the factor $D_{s\eta}(\bfk) = 1+e\boldsymbol{B}_\eta \cdot \boldsymbol{\Omega}_{s\eta}(\bfk)$ accounts for the modification of the phase space volume.
The net current, the spin current, and the valley current can be obtained by combining those $\{\boldsymbol{j}_{s\eta}\}$.
We estimate the current up to the first order in $\boldsymbol{B}_5$ and the second order in $\boldsymbol{E}_0$;
details of the solution process are shown in the Supplemental Material.
We should note that the intrinsic spin Hall current linear in $\boldsymbol{E}$ is not included in this formulation,
since the locations of the DPs are not taken into account.
In the field theory description, it is described by the Chern--Simons (or Bardeen--Zumino) terms \cite{Gorbar_2017a,Gorbar_2017b}.
However, since we are primarily interested in the nonequilibrium current in response to the electric field,
we first ignore it and later supplement it in the final discussion.

\

\textbf{Acknowledgments.} ---
This work is supported by JSPS KAKENHI Grant Number JP17K14316.
The author acknowledges K.~Kobayashi, K.~Nomura, and Y.~Ominato for fruitful discussions.

\textbf{Competing financial interests.} ---
The author declares no competing financial interests.



\begin{thebibliography}{99}
\bibitem{note1}
Note that this $\bfk$ corresponds to $\bfk - \eta K \hat{\boldsymbol{e}}_z$ in Eq.~(\ref{eq:Hamiltonian}).
\end{thebibliography}
\end{document}


\title{Supplemental Information for \\ ``Strain-induced nonlinear spin Hall effect in topological Dirac semimetal''}
\author{Yasufumi Araki}
\affiliation{Institute for Materials Research, Tohoku University, Sendai 980-8577, Japan}
\affiliation{Frontier Research Institute for Interdisciplinary Sciences, Tohoku University, Sendai 980-8578, Japan}

\preprint{Ver.~0.30}

\maketitle

\appendix

In this Supplemental Information, we show the details of our calculation,
especially the solution process of the Boltzmann equation in response to the electric field.
We here employ the perturbative expansion by $\boldsymbol{B}_5$ and neglect the terms beyond $O(B_5^1)$
in all the equations shown below.

\section{Fourier decomposition of the Boltzmann equation}
In order to estimate the response to the alternating electric field $\boldsymbol{E}(t) = \boldsymbol{E}_0 (e^{i\omega t}+ e^{-i \omega t})$,
we here decompose
the single-particle equations of motions
\begin{align}
\dot{\bfr} &= \boldsymbol{\nabla}_{\bfk} \tilde{\epsilon}_{s\eta}(\bfk) - \dot{\bfk} \times \boldsymbol{\Omega}_{s\eta}(\bfk) \label{eq:eom-r} \\
\dot{\bfk} &= -e\boldsymbol{E}_\eta - e\dot{\bfr} \times \boldsymbol{B}_{\eta} \label{eq:eom-k}
\end{align}
and the Boltzmann equation
\begin{align}
\left[ \dot{\bfr}\cdot\boldsymbol{\nabla}_{\bfr} + \dot{\bfk}\cdot\boldsymbol{\nabla}_{\bfk} + \partial_t \right]f_{s\eta}(\bfr,\bfk,t) = (df_{s\eta}/dt)_\mathrm{coll} \label{eq:Boltzmann}
\end{align}
into the Fourier components with the frequencies $0, \ \pm \omega,\ \pm 2\omega, \cdots$.

By solving Eqs.~(\ref{eq:eom-r}) and (\ref{eq:eom-k}) for $\dot{\bfr}$ and $\dot{\bfk}$,
we obtain
\begin{align}
D_{s\eta}\dot{\boldsymbol{r}} &= \tilde{\boldsymbol{v}}_{s\eta} + e\boldsymbol{E}_\eta \times \boldsymbol{\Omega}_{s\eta} +e(\tilde{\boldsymbol{v}}_{s\eta} \cdot\boldsymbol{\Omega}_{s\eta})\boldsymbol{B}_\eta \label{eq:eom-Dr} \\
D_{s\eta}\dot{\bfk} &= -e\boldsymbol{E}_\eta - e\tilde{\boldsymbol{v}}_{s\eta} \times\boldsymbol{B}_\eta - e^2 (\boldsymbol{E}_\eta \cdot \boldsymbol{B}_\eta) \boldsymbol{\Omega}_{s\eta}, \label{eq:eom-Dk}
\end{align}
where $\tilde{\boldsymbol{v}}_{s\eta}(\bfk) \equiv \boldsymbol{\nabla}_{\bfk} \tilde{\epsilon}_{s\eta}(\bfk)$
denotes the group velocity modified by the orbital magnetization.
The momentum-space Berry curvature $\boldsymbol{\Omega}_{s\eta}(\bfk)$
and the orbital magnetic moment $\boldsymbol{m}_{s\eta}(\bfk)$
for the eigenstate $u_{s\eta}(\bfk)$ are defined by
\begin{align}
\boldsymbol{\Omega}_{s\eta}(\bfk) &= -\im \left\langle \boldsymbol{\nabla}_{\bfk} u_{s\eta}(\bfk)| \times | \boldsymbol{\nabla}_{\bfk} u_{s\eta}(\bfk) \right\rangle \label{eq:Berry-curvature} \\
\boldsymbol{m}_{s\eta}(\bfk) &= -\frac{e}{2} \im \left\langle \boldsymbol{\nabla}_{\bfk} u_{s\eta}| \times \left[H_{s\eta}(\bfk) - \epsilon(\bfk) \right] | \boldsymbol{\nabla}_{\bfk} u_{s\eta} \right\rangle, \label{eq:orbital-magnetization}
\end{align}
respectively.
As the electromagnetic fields here read
\begin{align}
\boldsymbol{E}_\eta(t) = \boldsymbol{E}_0(e^{i\omega t}+e^{-i\omega t}), \quad \boldsymbol{B}_\eta = \eta \boldsymbol{B}_5,
\end{align}
the right-hand sides of Eqs. (\ref{eq:eom-Dr}) and (\ref{eq:eom-Dk})
can be decomposed into Fourier components as
\begin{align}
D_{s\eta} \dot{\bfr} &= \sum_{\nu=0,\pm 1} e^{i \nu\omega t} \boldsymbol{R}_{s\eta}^{(\nu \omega)} \\
D_{s\eta} \dot{\bfk} &= \sum_{\nu=0,\pm 1} e^{i \nu\omega t} \boldsymbol{K}_{s\eta}^{(\nu \omega)},
\end{align}
with
\begin{align}
\boldsymbol{R}_{s\eta}^{(0)} &= \tilde{\boldsymbol{v}}_{s\eta} +e(\tilde{\boldsymbol{v}}_{s\eta} \cdot\boldsymbol{\Omega}_{s\eta})\boldsymbol{B}_\eta \\
\boldsymbol{R}_{s\eta}^{(\pm \omega)} &= e \boldsymbol{E}_0 \times \boldsymbol{\Omega}_{s\eta} \label{eq:eom-R}\\
\boldsymbol{K}_{s\eta}^{(0)} &= - e\tilde{\boldsymbol{v}}_{s\eta} \times\boldsymbol{B}_\eta \\
\boldsymbol{K}_{s\eta}^{(\pm \omega)} &= -e\boldsymbol{E}_0 -e^2 (\boldsymbol{E}_0 \cdot \boldsymbol{B}_\eta) \boldsymbol{\Omega}_{s\eta} . \label{eq:eom-K}
\end{align}
Here the superscript in $(\cdot)$ on each physical quantity denotes its oscillation frequency.

Similarly, we also apply the Fourier decomposition
to the time-dependent distribution function $f_{s\eta}(\bfk,t)$,
which is given by
\begin{align}
f_{s\eta}(\bfk,t) = \sum_{n \in \mathbb{Z}} e^{i n\omega t} f_{s\eta}^{(n \omega)}(\bfk).
\end{align}
By using these Fourier components,
the Boltzmann equation
\begin{align}
\left[\dot{\bfk} \cdot \boldsymbol{\nabla}_{\bfk} +\partial_t \right] f_{s\eta}(\bfk,t) = - \tau^{-1} \left[f_{s\eta}(\bfk,t) - f_{s\eta}^\mathrm{eq}(\bfk) \right],
\end{align}
which is reduced from Eq.~(\ref{eq:Boltzmann}) by the homogeneous and relaxation-time approximations,
can be decomposed as
\begin{align}
& \left[\left(\boldsymbol{K}_{s\eta}^{(0)} \cdot \boldsymbol{\nabla}_{\bfk}\right) + D_{s\eta} \tau_n^{-1} \right] f_{s\eta}^{(n\omega)} - D_{s\eta} \tau^{-1} f_{s\eta}^\mathrm{eq} \delta_{n0} \label{eq:Boltzmann-iterative} \\
& + \left(\boldsymbol{K}_{s\eta}^{(+\omega)} \cdot \boldsymbol{\nabla}_{\bfk}\right) f_{s\eta}^{(n^- \omega)} + \left(\boldsymbol{K}_{s\eta}^{(-\omega)} \cdot \boldsymbol{\nabla}_{\bfk}\right) f_{s\eta}^{(n^+ \omega)} =0, \nonumber
\end{align}
with $\tau_n^{-1} \equiv \tau^{-1} + i n \omega$ and $n^\pm \equiv n \pm 1$,
for each frequency $n\omega$.
Each equation correlates three neighboring Fourier components of the distribution function;
by solving these equations iteratively, we can estimate the distribution function at any frequency.
Since $\boldsymbol{K}_{s\eta}^{(0)}$ is independent of $\boldsymbol{E}_0$ and $\boldsymbol{K}_{s\eta}^{(\pm \omega)}$ is linear in $\boldsymbol{E}_0$,
we can easily see from Eq.~(\ref{eq:Boltzmann-iterative}) that $f_{s\eta}^{(n\omega)}$ is of $O(E_0^{|n|})$
for small $E_0$.

Once we obtain the solution for those equations,
we can evaluate the Fourier components of the current for each spin and valley
as
\begin{align}
\boldsymbol{j}_{s\eta}^{(n\omega)} &= -e \int \frac{d^3\bfk}{(2\pi)^3} \left[ \boldsymbol{R}_{s\eta}^{(0)} f_{s\eta}^{(n\omega)} + \sum_{\pm}\boldsymbol{R}_{s\eta}^{(\pm \omega)} f_{s\eta}^{(n^\mp \omega)} \right],
\end{align}
which is based on the relation
\begin{align}
\boldsymbol{j}_{s\eta}(t) = -e \int \frac{d^3 \bfk}{(2\pi)^3} D_{s\eta}(\bfk) \dot{\bfr} f_{s\eta}(\bfk,t). \label{eq:current}
\end{align}
In particular, the stationary (DC) current can be given as
\begin{align}
\boldsymbol{j}_{s\eta}^{(0)} &= -e \int \frac{d^3\bfk}{(2\pi)^3} \left[ \boldsymbol{R}_{s\eta}^{(0)} f_{s\eta}^{(0)} + \sum_{\pm}\boldsymbol{R}_{s\eta}^{(\pm \omega)} f_{s\eta}^{(\mp \omega)} \right], \label{eq:j0}
\end{align}
which we shall use later to estimate the rectified current.

\section{Useful formulae in polar coordinate}
Before solving the Boltzmann equation iteratively,
we first list up some formula useful in solving the Boltzmann equation under the Dirac Hamiltonian.
Here we take the terms
up to the second order in $\boldsymbol{E}_0$ and the linear order in $\boldsymbol{B}_5$.

We mainly use the polar coordinate $(k,\theta,\phi)$ in the momentum space, defined by
\begin{align}
\boldsymbol{k} = k\left(\hat{\boldsymbol{e}}_x \sin\theta\cos\phi + \hat{\boldsymbol{e}}_y \sin\theta\sin\phi + \hat{\boldsymbol{e}}_z \cos\theta \right).
\end{align}
The basis vectors in this coordinate are defined by
\begin{align}
\hat{\bfk} &= \hat{\boldsymbol{e}}_x \sin\theta\cos\phi + \hat{\boldsymbol{e}}_y \sin\theta\sin\phi + \hat{\boldsymbol{e}}_z \cos\theta \\
\hat{\boldsymbol{e}}_\theta &= \hat{\boldsymbol{e}}_x \cos\theta\cos\phi + \hat{\boldsymbol{e}}_y \cos\theta\sin\phi - \hat{\boldsymbol{e}}_z \sin\theta \\
\hat{\boldsymbol{e}}_\phi &= -\hat{\boldsymbol{e}}_x \sin\phi + \hat{\boldsymbol{e}}_y \cos\phi,
\end{align}
and the gradient operator is given by
\begin{align}
\boldsymbol{\nabla}_{\bfk} = \hat{\bfk} \partial_k + \hat{\boldsymbol{e}}_\theta \frac{1}{k} \partial_\theta + \hat{\boldsymbol{e}}_\phi \frac{1}{k \sin\theta} \partial_\phi.
\end{align}

We should note the following formulae for gradient,
\begin{align}
\boldsymbol{\nabla}_{\bfk} k &= \hat{\bfk} \\
\boldsymbol{\nabla}_{\bfk} a_\parallel &= \frac{1}{k} \boldsymbol{a}_\perp
\end{align}
with $\boldsymbol{a}$ independent of $\bfk$.
Here $a_\parallel$ and $\boldsymbol{a}_\perp$ are decomposition of $\boldsymbol{a}$
into the components parallel and perpendicular to $\bfk$, respectively,
which are explicitly defined by
\begin{align}
a_\parallel \equiv \boldsymbol{a} \cdot \hat{\bfk}, \quad \boldsymbol{a}_\perp \equiv \boldsymbol{a} - a_\parallel \hat{\bfk}.
\end{align}

We also list up several formulae for the angular integral $\int d\Omega = \int d\theta \ \sin\theta \ d\phi$,
\begin{align}
\int d\Omega \ \hat{\bfk} & = 0 \label{eq:angular-1} \\
\int d\Omega \ a_{\parallel}\hat{\bfk} & = \frac{4\pi}{3}\boldsymbol{a} \label{eq:angular-2} \\
\int d\Omega \ a_{\parallel} b_\parallel \hat{\bfk} & = 0 \label{eq:angular-3}  \\
\int d\Omega \ a_{\parallel}^2 b_\parallel \hat{\bfk} & = \frac{4\pi}{15}\left[ 3(\boldsymbol{a}\cdot\boldsymbol{b})\boldsymbol{a} - (\boldsymbol{b}\times\boldsymbol{a})\times\boldsymbol{a}\right]. \label{eq:angular-4} 
\end{align}
The radial integral of the delta functions can be evaluated as
\begin{align}
\int_0^\infty dk \ k^n \delta(k-k_\mathrm{F}) &= k_\mathrm{F}^n \\
\int_0^\infty dk \ k^n \delta'(k-k_\mathrm{F}) &= - \int_0^\infty dk \ n k^{n-1} \delta(k-k_\mathrm{F}) \\
 &= -n k_\mathrm{F}^{n-1}.
\end{align}

\subsection{Explicit forms for Dirac/Weyl dispersion}
For a Weyl electron in the electron band,
the Berry curvature and the orbital magnetic moment, defined by Eqs.~(\ref{eq:Berry-curvature}) and (\ref{eq:orbital-magnetization}), are given by
\begin{align}
\boldsymbol{\Omega}_{s\eta}(\bfk) = s\eta \frac{1}{2k^2} \hat{\bfk}, \quad \boldsymbol{m}_{s\eta}(\bfk) = s\eta \frac{ev_\mathrm{F}}{2k}\hat{\bfk}.
\end{align}
Thus the phase space volume $D_{s\eta}$ and the modified single-particle energy $\tilde{\epsilon}_{s\eta}$ reduce to $\eta$-independent forms,
\begin{align}
D_{s\eta}(\bfk) &= 1 + e\boldsymbol{B}_\eta \cdot \boldsymbol{\Omega}_{s\eta}(\bfk) = 1 + s\frac{e B_{5\parallel}}{2k^2} \\
\tilde{\epsilon}_{s\eta}(\bfk) &= \epsilon(\bfk) - \boldsymbol{m}_{s\eta}(\bfk) \cdot \boldsymbol{B}_\eta \\
 &= v_\mathrm{F} k -s\frac{e v_\mathrm{F}}{2k} B_{5\parallel}.
\end{align}
The modified group velocity $\tilde{\boldsymbol{v}}_{s\eta}$ also becomes $\eta$-independent,
given as
\begin{align}
\tilde{\boldsymbol{v}}_{s\eta}(\bfk) &= \boldsymbol{\nabla}_{\boldsymbol{k}} \tilde{\epsilon}_{s\eta}(\bfk) \\
 &= v_\mathrm{F} \hat{\boldsymbol{k}} + s\frac{e v_\mathrm{F}}{2k^2} B_{5\parallel} \hat{\boldsymbol{k}} - s\frac{e v_\mathrm{F}}{2k^2} \boldsymbol{B}_{5\perp} \\
 &= \left(1 + s\frac{e B_{5\parallel}}{k^2} \right) v_\mathrm{F} \hat{\bfk} - s\frac{e v_\mathrm{F}}{2k^2} \boldsymbol{B}_5.
\end{align}
The equilibrium distribution under the axial magnetic field $\boldsymbol{B}_5$ is given as
\begin{align}
f_{s\eta}^\mathrm{eq}(\bfk) &= \theta\left(\mu - \tilde{\epsilon}_{s\eta}(\bfk)\right) \\
 &= \theta \left(k_\mathrm{F} - k + s\frac{e}{2k}B_{5\parallel} \right) \\
 & \simeq \theta(k_\mathrm{F}-k) + s\frac{e}{2k}B_{5\parallel} \delta(k-k_\mathrm{F}),
\end{align}
with the Fermi momentum $k_\mathrm{F} = \mu/v_\mathrm{F}$,
which is not perfectly symmetric around the DP due to the orbital magnetization.

$\boldsymbol{R}_{s\eta}^{(\nu \omega)}$ and $\boldsymbol{K}_{s\eta}^{(\nu \omega)}$ defined by Eqs.~(\ref{eq:eom-R}) and (\ref{eq:eom-K}) read
\begin{align}
\boldsymbol{R}_{s\eta}^{(0)} &= \left( 1+s\frac{e B_{5\parallel}}{k^2} \right) v_\mathrm{F} \hat{\bfk} \\
\boldsymbol{R}_{s\eta}^{(\pm\omega)} &= s\eta e \frac{1}{2k^2} \boldsymbol{E}_0 \times \hat{\bfk} \\
\boldsymbol{K}_{s\eta}^{(0)} &= -\eta e v_\mathrm{F} \hat{\bfk} \times \boldsymbol{B}_5 \\
\boldsymbol{K}_{s\eta}^{(\pm\omega)} &= -e \boldsymbol{E}_0 -s e^2\frac{\boldsymbol{E}_0 \cdot \boldsymbol{B}_5}{2k^2} \hat{\bfk},
\end{align}
up to $O(B_5^1)$.

\subsection{Linear response}

We first derive the linear response to the electric field,
which is represented by $f_{s\eta}^{(\pm\omega)}$.
It can be obtained by solving Eq.~(\ref{eq:Boltzmann-iterative}) for $n=\pm 1$,
which can be explicitly written as
\begin{align}
\left[\left(\boldsymbol{K}_{s\eta}^{(0)} \cdot \boldsymbol{\nabla}_{\bfk} \right) + D_{s\eta} \tau_\pm^{-1} \right]f_{s\eta}^{(\pm\omega)} = -\left(\boldsymbol{K}_{s\eta}^{(\pm\omega)} \cdot \boldsymbol{\nabla}_{\bfk} \right) f_{s\eta}^{\mathrm{eq}},
\end{align}
up to the linear order in $\boldsymbol{E}_0$.
Moving the first term on the left-hand side to the right-hand side,
this equation can be rewritten as
\begin{align}
f_{s\eta}^{(\pm\omega)} = F_{s\eta}^{(\pm\omega)\mathrm{a}} + F_{s\eta}^{(\pm\omega)\mathrm{b}} \label{eq:f-linear-iterative}
\end{align}
with
\begin{align}
F_{s\eta}^{(\pm\omega)\mathrm{a}} &= - D_{s\eta}^{-1} \tau_\pm \left(\boldsymbol{K}_{s\eta}^{(\pm\omega)} \cdot \boldsymbol{\nabla}_{\bfk} \right) f_{s\eta}^{\mathrm{eq}} \label{eq:f1-a} \\
F_{s\eta}^{(\pm\omega)\mathrm{b}} &= - D_{s\eta}^{-1} \tau_\pm \left(\boldsymbol{K}_{s\eta}^{(0)} \cdot \boldsymbol{\nabla}_{\bfk} \right) f_{s\eta}^{(\pm\omega)}. \label{eq:f1-b}
\end{align}
which can be evaluated iteratively by substituting $f_{s\eta}^{(\pm\omega)}$ obtained on the left-hand side of Eq.~(\ref{eq:f-linear-iterative})
to the right-hand side of Eq.~(\ref{eq:f1-b}).
Since $\boldsymbol{K}_{s\eta}^{(0)}$ is linear in $\boldsymbol{B}_5$,
each iteration yields a term higher by $B_5^1$.
We should make a single iteration to reach $f_{s\eta}^{(\pm\omega)}$ up to $O(B_5^1)$.

The first term $F_{s\eta}^{(\pm\omega)\mathrm{a}}$ can be given by evaluating
\begin{align}
F_{s\eta}^{(\pm\omega)\mathrm{a}} & \simeq \tau_\pm \left(1 -\frac{s e}{2k^2}B_{5\parallel} \right) \\
& \quad \quad \quad \times \left[e \boldsymbol{E}_0 +\frac{s e^2}{2k^2}(\boldsymbol{E}_0 \cdot \boldsymbol{B}_5) \hat{\bfk}\right] \cdot \boldsymbol{\nabla}_{\bfk} f_\eta^\mathrm{eq}, \nonumber
\end{align}
where we drop the terms higher than $O(B_5^1)$.
The gradient of $f_{s\eta}^\mathrm{eq}$ reads
\begin{align}
& \boldsymbol{\nabla}_{\bfk} f_\eta^\mathrm{eq} = \boldsymbol{\nabla}_{\bfk} \left[\theta(k_\mathrm{F}-k) + s\frac{e}{2k}B_{5\parallel} \delta(k-k_\mathrm{F})\right] \\
&\quad = \left[ -\hat{\bfk} + \frac{se}{2k^2}\boldsymbol{B}_5 - \frac{se}{k^2}B_{5\parallel} \hat{\bfk} + \frac{se}{2k}B_{5\parallel}\hat{\bfk} \partial_k \right] \delta_\mathrm{F}, \nonumber
\end{align}
where we employ shorthand notations
\begin{align}
\delta_\mathrm{F} \equiv \delta(k-k_\mathrm{F}), \quad \delta'_\mathrm{F} \equiv \partial_k \delta(k-k_\mathrm{F}), \quad \cdots.
\end{align}
By collecting the terms up to $O(B_5^1)$, we obtain
\begin{align}
F_{s\eta}^{(\pm\omega)\mathrm{a}} &= -e \tau_\pm E_{0\parallel}\delta_\mathrm{F} -\frac{se^2 \tau_n}{2k^2} B_{5\parallel} E_{0\parallel}(\delta_\mathrm{F}-k\delta'_\mathrm{F}) \label{eq:f1-1}
\end{align}

Since $\boldsymbol{K}_\eta^{(0)}$ is of $O(B_5^1)$,
$f_{s\eta}^{(\pm\omega)}$ at $O(B_5^0)$ can be simply given as
\begin{align}
f_{s\eta}^{(\pm\omega)} = -e \tau_\pm E_{0\parallel} \delta(k-k_\mathrm{F}) + O(B_5^1)
\end{align}
arising from $F_{s\eta}^{(\pm\omega)\mathrm{a}}$.
Therefore, $F_{s\eta}^{(\pm\omega)\mathrm{b}}$ up to $O(B_5^1)$ can be obtained from Eq.~(\ref{eq:f1-b}) as
\begin{align}
F_{s\eta}^{(\pm\omega)\mathrm{b}} &\simeq -D_{s\eta}^{-1} \tau_\pm \left(\boldsymbol{K}_{s\eta}^{(0)} \cdot \boldsymbol{\nabla}_{\bfk} \right) F_{s\eta}^{(\pm\omega)\mathrm{a}} \bigr|_{\boldsymbol{B}_5=0} \\
&\simeq -\eta e^2 v_\mathrm{F} \tau_\pm^2 \left(\hat{\bfk}\times\boldsymbol{B}_5\right) \cdot \boldsymbol{\nabla}_{\bfk} \left[E_{0\parallel} \delta_\mathrm{F}\right] \\
&= -\eta e^2 v_\mathrm{F} \tau_\pm^2 \left(\hat{\bfk}\times\boldsymbol{B}_5\right) \cdot \left[\frac{1}{k} \boldsymbol{E}_{0\perp} \delta_\mathrm{F} + E_{0\parallel} \delta'_\mathrm{F} \hat{\bfk} \right] \\
&= -\eta e^2 v_\mathrm{F} \tau_\pm^2 \left(\hat{\bfk}\times\boldsymbol{B}_5\right) \cdot \boldsymbol{E}_0 \frac{1}{k} \delta_\mathrm{F} \\
&= -\eta e^2 v_\mathrm{F} \tau_\pm^2 \frac{1}{k} \left(\boldsymbol{B}_5 \times \boldsymbol{E}_0\right)_\parallel \delta_\mathrm{F}, \label{eq:f1-2}
\end{align}
where one should note that $\hat{\bfk} \times \boldsymbol{B}_5$ is perpendicular to $\hat{\bfk}$.

By collecting Eqs.~(\ref{eq:f1-1}) and (\ref{eq:f1-2}),
$f_\eta^{(\pm\omega)}(\bfk)$ reads
\begin{align}
f_{s\eta}^{(\pm \omega)}(\bfk) &= - e \tau_\pm E_{0\parallel} \delta_\mathrm{F} -\eta \frac{e^2 v_\mathrm{F} \tau_\pm^2}{k}(\boldsymbol{B}_5 \times \boldsymbol{E}_0)_\parallel \delta_\mathrm{F} \label{eq:f1} \\
& \quad \quad -s\frac{e^2 \tau_\pm}{2k^2} E_{0\parallel} B_{5\parallel} (\delta_\mathrm{F}-k\delta'_\mathrm{F}), \nonumber
\end{align}
up to $O(E_0^1, B_5^1)$.
The first term in $[\cdots]$ represents the shift of the distribution from the drift by $\boldsymbol{E}(t)$,
while the second term gives the shift from the regular Hall effect by the \textit{axial} magnetic field $\boldsymbol{B}_5$ and the \textit{normal} electric field $\boldsymbol{E}(t)$.
The third term has a complicated structure:
it gives a \textit{quadrupolar} distribution in $\bfk$-space unless $\boldsymbol{E}_0$ and $\boldsymbol{B}_5$ are exactly parallel.
It arises as the modification to the first (drift) term from the Berry curvature and the orbital magnetic moment.
This term represents the effect of the chiral anomaly:
it modifies the electron density, whose deviation from equilibrium is given by
\begin{align}
& \delta \rho_{s\eta}(t) = \int \frac{d^3 \bfk}{(2\pi)^3} D_{s\eta}(\bfk) \sum_{\pm} e^{\pm i\omega t} f_{s\eta}^{(\pm\omega)}(\bfk) \\
& \quad -\frac{s e^2}{4\pi^2} \frac{2\tau}{1+\omega^2 \tau^2} \boldsymbol{E}_0 \cdot \boldsymbol{B}_5 (\cos\omega t +\omega \tau \sin\omega t),
\end{align}
where we have used
\begin{align}
\int \frac{d^3 \bfk}{(2\pi)^3} D_{s\eta}(\bfk) f_{s\eta}^{(\pm\omega)}(\bfk) = -s \frac{e^2 \tau_\pm}{4\pi^2} \boldsymbol{E}_0 \cdot \boldsymbol{B}_5.
\end{align}
We can see that the induced electron density $\delta \rho_{s\eta}(t)$ for each valley/spin satisfies the continuity equation
\begin{align}
\left[\partial_t +\frac{1}{\tau} \right] \delta \rho_{s\eta}(t) = -s\frac{e^2}{4\pi^2}\boldsymbol{B}_5 \cdot \boldsymbol{E}(t), \label{eq:anomaly}
\end{align}
modified by the chiral anomaly,
noting that $\boldsymbol{E}(t) = 2\boldsymbol{E}_0 \cos\omega t$.
Although there is no spin flip term in the original Hamiltonian (except for the $\bfk^3$ terms),
the anomaly violates spin conservation if $\boldsymbol{B}_5 \cdot \boldsymbol{E}_0 \neq 0$,
leading to the spin polarization oscillating with the frequency $\omega$.

\subsection{Quadratic response}
Next we solve Eq.~(\ref{eq:Boltzmann-iterative}) for $n=0$
to evaluate the stationary distribution $f_{s\eta}^{(0)}$ up to the second order in $\boldsymbol{E}_0$.
The equation can be explicitly written as
\begin{align}
\left[\left(\boldsymbol{K}_{s\eta}^{(0)} \cdot \boldsymbol{\nabla}_{\bfk} \right) + D_{s\eta} \tau^{-1} \right] \delta f_{s\eta}^{(0)}
 = - \sum_{\pm} \left(\boldsymbol{K}_{s\eta}^{(\mp\omega)} \cdot \boldsymbol{\nabla}_{\bfk} \right)f_{s\eta}^{(\pm\omega)},
\end{align}
where $\delta f_{s\eta}^{(0)} \equiv f_{s\eta}^{(0)} - f_{s\eta}^\mathrm{eq}$.
Similarly to Eq.~(\ref{eq:f-linear-iterative}), this equation can be rewritten as
\begin{align}
\delta f_{s\eta}^{(0)} = F_{s\eta}^{(0)\mathrm{a}} + F_{s\eta}^{(0)\mathrm{b}} \label{eq:f-quadratic-iterative}
\end{align}
with
\begin{align}
F_{s\eta}^{(0)\mathrm{a}} &= -D_{s\eta}^{-1} \tau \sum_{\pm} \left(\boldsymbol{K}_{s\eta}^{(\mp\omega)} \cdot \boldsymbol{\nabla}_{\bfk} \right)f_{s\eta}^{(\pm\omega)} \label{eq:f0-a} \\
F_{s\eta}^{(0)\mathrm{b}} &= -D_{s\eta}^{-1} \tau \left(\boldsymbol{K}_{s\eta}^{(0)} \cdot \boldsymbol{\nabla}_{\bfk} \right)\delta f_{s\eta}^{(0)}, \label{eq:f0-b}
\end{align}
which can again be evaluated iteratively.

Evaluation of $F_{s\eta}^{(0)\mathrm{a}}$ is lengthy but straightforward.
The gradient of $f_{s\eta}^{(\pm\omega)}$ is given as
\begin{align}
& \boldsymbol{\nabla}_{\bfk} f_{s\eta}^{(\pm\omega)} = -e \tau_n \left(\frac{1}{k} \boldsymbol{E}_{0\perp} \delta_\mathrm{F} + \hat{\bfk} \delta'_\mathrm{F} \right)  \\
& \quad + \frac{se^2 \tau_n}{k^3} B_{5\parallel} E_{0\parallel} \hat{\bfk} \left( \delta_\mathrm{F} -k\delta'_\mathrm{F} +\frac{k^2}{2}\delta''_\mathrm{F} \right) \nonumber \\
& \quad - \frac{se^2 \tau_n}{2k^3} \left(\boldsymbol{B}_{5\perp} E_{0\parallel} + B_{5\parallel} \boldsymbol{E}_{0\perp}\right) (\delta_\mathrm{F} -k\delta'_\mathrm{F}) \nonumber \\
& \quad -\frac{\eta e^2 v_\mathrm{F} \tau_n^2}{k^2} \left[ (\boldsymbol{B}_5 \times \boldsymbol{E}_0) \delta_\mathrm{F} +(\boldsymbol{B}_5 \times \boldsymbol{E}_0)_\parallel (-2 \delta_\mathrm{F} + k \delta'_\mathrm{F}) \right]. \nonumber
\end{align}
Substituting this form into Eq.~(\ref{eq:f0-a}) and collecting the terms up to $O(B_5^1)$,
we obtain
\begin{align}
F_{s\eta}^{(0)\mathrm{a}} &= -\frac{2e^2 \tau \tau_*}{k} \left[ (E_0^2 -E_{0\parallel}^2)\delta_\mathrm{F} + E_{0\parallel}^2 k \delta'_\mathrm{F} \right] \label{eq:f0-1} \\
& \quad +\frac{s e^3 \tau \tau_*}{k^3} \Bigl[ -(\boldsymbol{E}_0 \cdot \boldsymbol{B}_5)E_{0\parallel} \delta_\mathrm{F} +E_0^2 B_{0\parallel} k \delta'_\mathrm{F} \nonumber \\
& \quad \quad \quad \quad \quad \quad +E_{0\parallel}^2 B_{5\parallel} (3\delta_\mathrm{F} - 3k \delta'_\mathrm{F} +2k^2 \delta''_\mathrm{F}) \Bigr] \nonumber \\
& \quad + \frac{2\eta e^3 v_\mathrm{F} \tau \tau_{**}^2}{k^2}(\boldsymbol{B}_5 \times \boldsymbol{E}_0)_\parallel E_{0\parallel}(2 \delta_\mathrm{F} - k \delta'_\mathrm{F}), \nonumber
\end{align}
where $\tau_*$ and $\tau_{**}$ are defined as
\begin{align}
\tau_* &= \frac{1}{2}\left(\tau_{+} + \tau_{-} \right) = \tau \frac{1}{1+\omega^2 \tau^2} \\
\tau_{**}^2 &= \frac{1}{2}\left(\tau_{+}^2 + \tau_{-}^2 \right) = \tau^2 \frac{1-\omega^2 \tau^2}{(1+\omega^2 \tau^2)^2}.
\end{align}

Since $\boldsymbol{K}_\eta^{(0)}$ is of $O(B_5^1)$,
$\delta f_{s\eta}^{(0)}$ at $O(B_5^0)$ is given by the first line in Eq.~(\ref{eq:f0-1}).
Therefore, by substituting this part to Eq.~(\ref{eq:f0-b}),
we obtain
\begin{align}
F_{s\eta}^{(0)\mathrm{b}} &\simeq -D_{s\eta}^{-1} \tau \left(\boldsymbol{K}_{s\eta}^{(0)} \cdot \boldsymbol{\nabla}_{\bfk} \right) \delta F_{s\eta}^{(0)\mathrm{a}} \bigr|_{\boldsymbol{B}_5=0} \\
 & \simeq -2\eta e^3 v_\mathrm{F} \tau^2 \tau_* (\hat{\bfk} \times \boldsymbol{B}_5) \nonumber \\
 & \quad \quad \quad  \cdot \boldsymbol{\nabla}_{\bfk} \left[ (E_0^2 -E_{0\parallel}^2) k^{-1} \delta_\mathrm{F} + E_{0\parallel}^2 \delta'_\mathrm{F} \right] \nonumber \\
 & = -2\eta e^3 v_\mathrm{F} \tau^2 \tau_* (\hat{\bfk} \times \boldsymbol{B}_5) \cdot E_{0\parallel} \boldsymbol{E}_0 \left[ -\frac{2}{k^2}  \delta_\mathrm{F} + \frac{2}{k} \delta'_\mathrm{F} \right] \\
 & = \frac{4\eta e^3 v_\mathrm{F} \tau^2 \tau_*}{k^2} (\boldsymbol{B}_5 \times \boldsymbol{E}_0)_\parallel E_{0\parallel} (\delta_\mathrm{F} - k \delta'_\mathrm{F}). \label{eq:f0-2}
\end{align}
Collecting Eqs.~(\ref{eq:f0-1}) and (\ref{eq:f0-2}), we obtain the stationary response $\delta f_{s\eta}^{(0)}$.

\section{Stationary current}
Finally, we calculate the nonlinear stationary current from the distribution functions obtained above.
From Eq.~(\ref{eq:j0}), we split the stationary current $\boldsymbol{j}_{s\eta}^{(0)}$ into three parts,
\begin{align}
\boldsymbol{j}_{s\eta}^{(0)} = \boldsymbol{j}_{s\eta}^{(0)\mathrm{eq}} + \boldsymbol{j}_{s\eta}^{(0)\mathrm{a}} + \boldsymbol{j}_{s\eta}^{(0)\mathrm{b}},
\end{align}
with
\begin{align}
\boldsymbol{j}_{s\eta}^{(0)\mathrm{eq}} &= -\frac{e}{(2\pi)^3} \int d^3 \bfk \ \boldsymbol{R}_{s\eta}^{(0)}(\bfk) f_{s\eta}^\mathrm{eq}(\bfk) \\
\boldsymbol{j}_{s\eta}^{(0)\mathrm{a}} &= -\frac{e}{(2\pi)^3} \int d^3 \bfk \ \boldsymbol{R}_{s\eta}^{(0)}(\bfk) \delta f_{s\eta}^{(0)}(\bfk) \\
\boldsymbol{j}_{s\eta}^{(0)\mathrm{b}} &= -\frac{e}{(2\pi)^3} \int d^3 \bfk \ \sum_{\pm} \boldsymbol{R}_{s\eta}^{(\pm\omega)}(\bfk) f_{s\eta}^{(\mp\omega)}(\bfk).
\end{align}
Here $\boldsymbol{j}_{s\eta}^{(0)\mathrm{eq}}$ accounts for the equilibrium current that is present even in the absence of the electric field $\boldsymbol{E}(t)$,
whereas the other two are nonequilibrium current driven by the electric field;
$\boldsymbol{j}_{s\eta}^{(0)\mathrm{a}}$ comes from the stationary distribution $f_{s\eta}^{(0)}$ and the drift velocity $\boldsymbol{R}_{s\eta}^{(0)}$,
while $\boldsymbol{j}_{s\eta}^{(0)\mathrm{b}}$ is from the oscillating distribution $f_{s\eta}^{(\mp\omega)}$ and the anomalous velocity $\boldsymbol{R}_{s\eta}^{(\pm\omega)}$.

The equilibrium current $\boldsymbol{j}_{s\eta}^{(0)\mathrm{eq}}$ can be straightforwardly evaluated as
\begin{align}
\boldsymbol{j}_{s\eta}^{(0)\mathrm{eq}} &=  -\frac{e v_\mathrm{F}}{(2\pi)^3} \int d^3 \bfk \left(1 + \frac{se}{k^2} B_{5\parallel} \right) \\
 & \quad \quad \quad \quad \quad \times \left[ \theta(k_\mathrm{F}-k) + \frac{se}{2k} B_{5\parallel} \delta(k-k_\mathrm{F}) \right] \hat{\bfk} \nonumber \\
 &= -s \frac{e^2 v_\mathrm{F}}{4\pi} k_\mathrm{F} \boldsymbol{B}_5 = -s \frac{e^2 \mu}{4\pi} \boldsymbol{B}_5, \label{eq:j0-eq}
\end{align}
which accounts for the chiral axial (pseudo) magnetic effect under the axial magnetic field $\boldsymbol{B}_5$,
given by $\boldsymbol{J}^{(0)}_\mathrm{eq}$ in the main text.

$\boldsymbol{j}_{s\eta}^\mathrm{(0)a}$ can be obtained by evaluating the $\bfk$-space integral
\begin{align}
\boldsymbol{j}_{s\eta}^\mathrm{(0)a} &= -e \int \frac{d^3\bfk}{(2\pi)^3} \boldsymbol{R}_\mathrm{s\eta}^{(0)}(\bfk) \delta f_{s\eta}^{(0)}(\bfk) \\
 &= -ev_\mathrm{F} \int \frac{d^3\bfk}{(2\pi)^3} \hat{\bfk} \left(1+\frac{s e}{k^2} B_{5\parallel} \right)\delta f_{s\eta}^{(0)}(\bfk).
\end{align}
In the integrand function
\begin{align}
\left(1+\frac{s e}{k^2} B_{5\parallel}\right) \delta f_{s\eta}^{(0)}(\bfk),
\end{align}
the terms surviving after the angular integral can be extracted,
by using the angular integral formulae Eq.~(\ref{eq:angular-1})-(\ref{eq:angular-4}), as
\begin{align}
& \frac{s e^3 \tau \tau_*}{k^3} \Bigl[ -(\boldsymbol{E}_0 \cdot \boldsymbol{B}_5)E_{0\parallel} \delta_\mathrm{F} +E_0^2 B_{0\parallel} (-2 \delta_\mathrm{F} - k \delta'_\mathrm{F}) \nonumber \\
& \quad \quad \quad \quad \quad \quad +E_{0\parallel}^2 B_{5\parallel} (5\delta_\mathrm{F} - 5k \delta'_\mathrm{F} +2k^2 \delta''_\mathrm{F}) \Bigr].
\end{align}
By evaluating the angular and radial integrals, we obtain
\begin{align}
\boldsymbol{j}_{s\eta}^\mathrm{(0)a} &= \frac{s e^4 v_\mathrm{F} \tau \tau_*}{(2\pi)^3 k_\mathrm{F}} \biggl[ (\boldsymbol{E}_0 \cdot \boldsymbol{B}_5) \frac{4\pi}{3} \boldsymbol{E}_0 +2 E_0^2 \frac{4\pi}{3} \boldsymbol{B}_5 \\
 & \quad \quad -5\left(\frac{4\pi}{5}(\boldsymbol{E}_0 \cdot \boldsymbol{B}_5)\boldsymbol{E}_0 -\frac{4\pi}{15} (\boldsymbol{B}_5 \times \boldsymbol{E}_0) \times \boldsymbol{E}_0 \right) \biggr] \nonumber \\
 &= - s \frac{e^4 v_\mathrm{F} \tau \tau_*}{6\pi^2 k_\mathrm{F}} (\boldsymbol{B}_5 \times \boldsymbol{E}_0) \times \boldsymbol{E}_0, \label{eq:j0-a}
\end{align}
where we have used the vector triple product relation
\begin{align}
(\boldsymbol{B}_5 \times \boldsymbol{E}_0) \times \boldsymbol{E}_0 = (\boldsymbol{E}_0 \cdot \boldsymbol{B}_5)\boldsymbol{E}_0 - E_0^2 \boldsymbol{B}_5.
\end{align}

$\boldsymbol{j}_{s\eta}^\mathrm{(0)b}$ is given by the $\bfk$-space integral
\begin{align}
\boldsymbol{j}_{s\eta}^{(0)\mathrm{b}} &= -e \int \frac{d^3 \bfk}{(2\pi)^3} \ \sum_{\pm} \boldsymbol{R}_{s\eta}^{(\pm\omega)}(\bfk) f_{s\eta}^{(\mp\omega)}(\bfk) \\
 &= -s\eta e^2 \boldsymbol{E}_0 \times \int \frac{d^3 \bfk}{(2\pi)^3} \ \hat{\bfk} \frac{1}{2k^2} \sum_\pm f_{s\eta}^{(\mp\omega)}(\bfk).
\end{align}
In the integrand function
\begin{align}
\frac{1}{2k^2} \sum_\pm f_{s\eta}^{(\mp\omega)}(\bfk),
\end{align}
the term surviving after the angular integral can be extracted as
\begin{align}
-\eta e^2 v_\mathrm{F} \tau_{**}^2 \frac{1}{k^3} \left(\boldsymbol{B}_5 \times \boldsymbol{E}_0\right)_\parallel \delta_\mathrm{F}.
\end{align}
By evaluating the angular and radial integrals, we obtain
\begin{align}
\boldsymbol{j}_{s\eta}^{(0)\mathrm{b}} &= \frac{s e^4 v_\mathrm{F} \tau_{**}^2}{(2\pi)^3 k_\mathrm{F}} \boldsymbol{E}_0 \times \frac{4\pi}{3} \left(\boldsymbol{B}_5 \times \boldsymbol{E}_0\right) \\
 &= -\frac{s e^4 v_\mathrm{F} \tau_{**}^2}{6\pi^2 k_\mathrm{F}}\left(\boldsymbol{B}_5 \times \boldsymbol{E}_0\right) \times \boldsymbol{E}_0. \label{eq:j0-b}
\end{align}

By collecting Eqs.~(\ref{eq:j0-a}) and (\ref{eq:j0-b}),
we obtain the final form given in the main text.